\documentclass[a4paper,12pt]{article}

\begin{document}
\makeatletter
\renewcommand{\theequation}{\thesection.\arabic{equation}}
\@addtoreset{equation}{section}
\makeatother

\title{ An attempt to resolve the cosmological constant problem in the modified Yang's noncommutative quantized space-time }

\author{\large Sho Tanaka\footnote{Em. Professor of Kyoto University, E-mail: st-desc@kyoto.zaq.ne.jp }
\\[8 pt]
 Kurodani 33-4, Sakyo-ku, Kyoto 606-8331, Japan}
\date{}
\maketitle

\vspace{-10cm}
\rightline{}
\vspace{10cm}

\abstract{We attempt to resolve the cosmological constant problem through the key concept of the quantized number of spatial degrees of freedom in the modified Yang's quantized space-time, $n_{\rm {dof}} (V_3^{R(\tau)}).$

\vspace{0.5cm}
Keywords: the cosmological constant problem, modified Yang's noncommutative quantized space-time, quantized number of spatial degrees of freedom of our universe, the big bang stage, the arrow of cosmological time}

\section{\normalsize Introduction}

In our preceding papers, ``Where does black-hole entropy lie?" [1] and ``A short essay on quantum black holes and underlying noncommutative quantized space-time"[2] hereafter referred as I and II, respectively, we emphasized the importance of the underlying noncommutative geometry such as Snyder's and Yang`s noncommutative quantized space-time [3-6] towards the ultimate theory of quantum gravity or Planck scale physics. In the present paper, we more specifically notice the {\it modified} Yang's noncommutative quantized space-time. 

In fact, at the end of {\bf  5. Concluding arguments and further outlook} in II,  we stated: ``Before closing this short essay, let us note another interesting possibility of Yang's quantized space-time algebra (YSTA, see Appendix A). Indeed, one should notice that YSTA  is intrinsically equipped with the long scale parameter {\it R}, together with the short scale parameter {\it a}\footnote{ In our present researches, we have described so far the short scale parameter in YSTA by $\lambda$ instead of {\it a} which C.N.~Yang [5-6] used in accord with S.~Snyder [[3-4], as early  footnoted inside Appendix A in II. }
which has been identified with Planck length $ l_P $ in our present research so far. On the other hand, as was preliminarily pointed out in [25], ${\it R}$ might be promisingly related to a {\it fundamental cosmological constant} in connection with the recent dark-energy problem, under the further idea that YSTA subject to the {\it SO(D+1,1)} algebra might be understood in terms of  some kind of local reference frame in the ultimate theory of quantum gravity, on the analogy of the familiar local Lorentz frame in  Einstein's General Theory of Relativity."

The above statement must be just the subject of the present paper. In this connection, it might be appropriate here in advance to mention the {\it modified version} of  Yang's original quantized space-time (see Appendix A ). 

Indeed, as will be shown in Appendix A,  we intend in its modified version (MYST) to replace the long scale parameter $R$ in the original version (YST) to be some kind of ${\it time}\ \tau -$dependent: let us say it the {\it cosmological time} dependent, $R(\tau)$ ( see the beginning of section {\bf 3} ).  In fact, by this treatment, we intend that  the series of MYST characterized by $(\lambda, R(\tau))$ becomes able to describe the {\it historical development} of our universe beginning with the so-called big bang. 

\section{\normalsize The essence of cosmological constant problem}

Today, we know that the observed value of cosmological constant $\Lambda_{obs.}$  is over one hundred twenty orders of magnitude smaller  than the Planck energy density. That is, 

\begin{eqnarray}
  \Lambda_{\rm obs.} \sim 10^{-122} G^{-2},
\end{eqnarray}
where $G$ denotes the Newton constant and $G^{-2}$ is known to be the so-called Planck energy density, see for instance [7].

At this point, we are almost uniquely led to the following   {\it theoretical} expression $\Lambda (\tau)_{\rm theor.}$ at the present cosmological time $\tau$

\begin{eqnarray}
   \Lambda (\tau)_{\rm theor.} = {n_{\rm dof} ( V_3^{R(\tau)}) }^{-1} G^{-2}
\end{eqnarray}
on the basis of  $ n_{\rm dof} ( V_3^{R(\tau)}),$ which is the key concept in the modified Yang's quantized space-time (see Appendix A), and most importantly describes the quantized number of spatial degrees of freedom inside  $V_3^{R(\tau)}$, i.e., $d=3$ - dimensional volume with radius $R(\tau)$:
\begin{eqnarray}
{n_{\rm dof}} ( V_3^{R(\tau)} ) = ( [R(\tau)/\lambda] +1)^2, 
\end{eqnarray} 
( see (3.1) in II ), where $ [R(\tau)/\lambda] $ means the nearest integer of $ R(\tau)/\lambda. $

 In fact, one can easily confirm that  the factor ${n_{\rm dof} ( V_3^{R(\tau)}) }^{-1} ( \sim(\lambda/R(\tau))^2 $ in (2.2) just presents the extremely small factor
 $10^{-122}$ in (2.1) under $R(\tau) \sim 10^{28} {\rm cm}$ and $\lambda (= l_P) \sim 10^{-33} {\rm cm}.$

That is, the essence of cosmological constant problem is now clearly understood or resolved in the fundamental structure of $\Lambda (\tau)_{theor.} $ derived in (2.2) where the most  puzzling factor $10^{-122}$  in the  $\Lambda_{\rm obs.} $  in (2.1) is {\it theoretically} understood in terms of  the key concept ${n_{\rm dof}( V_3^{R(\tau)}) }.$

By the way,  $n_{\rm dof} ( V_3^{R(\tau)} )$  and the relation (2.3) itself as a whole were derived according to the general consideration of  ``Kinematical Holographic Relation" {\bf KHR} given by (3.1) in II, through the proper irreducible representation of Yang's quantized space-time algebra YSTA (see Appendix A in II ), while it is now developed on the modified Yang's quantized space-time algebra (MYSTA) (see Appendix A).

Furthermore, one should notice the fact that $n_{\rm dof} ( V_3^{R(\tau)} )$ is nothing but the whole number of  quantized spatial degrees of freedom inside of $ V_3^{R(\tau)},$ that is, the whole spatial volume of our universe at the cosmological time $\tau$, as seen from the consideration given in I,  II and originally in [8]. 

\section{\normalsize Concluding arguments and further outlook } 

First, le us notice that the theoretical cosmological constant $\Lambda (\tau)_{\rm theor.} = {n_{\rm dof} ( V_3^{R(\tau)}) }^{-1} G^{-2}$ derived in Eq. (2.2) becomes the cosmological time ${\tau}$ dependent through the term ${n_{\rm dof} ( V_3^{R(\tau)})}$, as noted in Appendix A in general. In this connection, it is quite important to notice that the term ${n_{\rm dof} ( V_3^{R(\tau)})}$ is directly related to the {\it entropy} of the static and equilibrium system in the following way

\begin{eqnarray}
 S(V_3^{R(\tau)})  ( = - {\rm Tr} [{\bf W} (V_3^{R(\tau)}){\rm ln} {\bf W} (V_3^{R(\tau)})] )
= n_{\rm dof} (V_3^{R(\tau)}) S[site] 
\nonumber\\
= ([R(\tau)/\lambda]+1)^2 S[site],
\end{eqnarray}
as seen in Eq.(3.10) in II with {\it L} replaced by $R(\tau)$. In the above expression $S[site]$ denotes the entropy assumed to be commonly realized in every [site], that is, every basis vector in Hilbert space II (see, section {\bf 2} in II). This fact guarantees that the cosmological time $ {\tau}$ acquires the so-called arrow of time, beyond a simple parameter (see Appendix {\bf A} ).

Second, let us briefly consider the big bang stage. We assume that it starts at $\tau \sim 0$  with the uncertainty $\lambda.$ Then, one finds in Eq.(2.2)  ${\rm n_{dof}} ( V_3^{\sim \lambda}) \sim 4 $ and
\begin{eqnarray} 
\Lambda (\lambda)_{\rm theor.} \sim {1\over 4} \ G^{-2}
\end{eqnarray}
at the very beginning stage of the big bang, in a sharp contrast to Eq. (2.1).  It reminds us the extremely microscopic black hole system considered in the subsection {\bf 4.3} in II.  It is our great interest further to investigate the later various stages of the big bang in accordance with the consideration of the black holes ranging over from macroscopic to extremely microscopic scales given also in {\bf  4.3} in II.

It should be furthermore noted here that the above consideration of the big bang might be fulfilled without suffering from the so-called ``Hawking-Penrose singularity theorems," because of the fact that our present research based on the noncommutative geometry or quantized space-time is well associated with the Heisenberg's Uncertainty Principle, so as to be free from singularities (see Appendix B in II, ``Historical background  of noncommutative quantized space and time").  

In this connection, it might be possibly claimed that the above derivation of $\Lambda (\tau)_{\rm theor.} = {n_{\rm dof} ( V_3^{R(\tau)}) }^{-1} G^{-2}$ in Eq.(2.2) seems a bit  {\it ad hoc}  and a certain numerical fine tuning to explain the puzzling factor $10^{-122}$  in the  Eq.(2.1) {\it accidentally} through $ n_{\rm dof} ( V_3^{R(\tau)}) .$  However, the present author should like to notice that on the back of  the key concept  $ n_{\rm dof}(V_3^{R(\tau)}),$  clearly lies the idea of  H.~ Yukawa's Atomism of quantized space-time or Elementary Domain beginning from the preceding idea of P.A.M.~Dirac's  Generalized transformation function (`g.t.f.' see Appendix B in II), which governs the whole space of universe, in assosiation with  the present cosmological constant problem or the dark energy problem. We anticipate that our $ \Lambda (\tau)_{\rm theor.} $ given by Eq. (2.2) will show widely the {\it predictive} power in the later various stages of the big bang over the present single prediction: $\Lambda (\lambda)_{\rm theor.} \sim {1\over 4} \ G^{-2} $ derived in Eq.(3.2) for its very beginning stage. 
    
In this article we have omitted the important arguments related to the so-called standard model of high energy physics. It is our important task to reconstruct the M -theory [9] in terms of the present Modified Yang's Lorentz covariant quantized space-time towards the ultimate theory of quantum gravity and quantum cosmology.

\vskip 1.5cm
\centerline {\Large  \bf Acknowledgments}
\vskip 0.8cm

The author would like to thank Taichiro Kugo for valuable discussions from the different point of view. The author is grateful to Hideaki Aoyama for giving constant encouragement to the present research.

\vskip 1.5cm

\centerline{\Large\bf Appendix }
\vskip 1cm
\appendix

\section{ Modified Yang's Lorentz covariant quantized space-time (MYST)}
\label{appendixa}

 Even in the MYST, as in the original Yang's quantized space-time (YST)  (see Appendix A in II) it is equipped with the so-called Inonu-Wigner's two contraction parameters, long $R$ and short $\lambda$.  The MYST is simply given  by replacing the long scale parameter $R$ in YST to be the {\it cosmological time} $\tau-$dependent  $R(\tau) $, as noted at the end of {\bf Introduction}.

In addition, we  intend now to introduce this cosmological time $\tau$ so as to describe the observation time of  $\Lambda_{obs.} ( \sim 10^{-122} G^{-2}$ ) in (2.1) on the one hand and on the other hand, the set of long and short scale parameters $ ( R(\tau), \lambda)$ at $\tau$  to specify the modified Yang's quantized space-time as the {\it local reference frame} at the cosmological time $\tau$, as was mentioned in {\bf Introduction}, according to the idea of the local Lorentz frame in General Theory of Relativity.

First,  (A.1) holds as it stands
\begin{eqnarray} 
 \hat{\Sigma}_{MN}  \equiv i (q_M \partial /{\partial{q_N}}-q_N\partial/{\partial{q_M}})
\label{sec_a: stp}
\end{eqnarray}
and  (A.2)  clearly tends into
\begin{eqnarray}
             - q_0^2 + q_1^2 + \cdots + q_{D-1}^2 + q_a^2 + q_b^2 = R (\tau)^2.
\end{eqnarray}

Now,  $D$-dimensional space-time and momentum operators, $\hat{X}_\mu$ and $\hat{P}_\mu$, 
with $\mu =1,2,\cdots,D,$ are defined by
\begin{eqnarray}
&&\hat{X}_\mu \equiv \lambda\ \hat{\Sigma}_{\mu a}
\\
&&\hat{P}_\mu (\tau) \equiv \hbar /R (\tau) \hat{\Sigma}_{\mu b},   
\end{eqnarray}
together with $D-$dimensional angular momentum $\hat{M}_{\mu \nu}$
\begin{eqnarray}
   \hat{M}_{\mu \nu} \equiv \hbar \hat{\Sigma}_{\mu \nu}
\end{eqnarray} 
and the so-called reciprocity operator
\begin{eqnarray}
    \hat{N} (\tau) \equiv \lambda /R (\tau) \hat{\Sigma}_{ab}.
\end{eqnarray}

Here and hereafter, one should notice that in MYST the momentum operator or the reciprocity operator and so on become the  cosmological constant $\tau-$dependent in general. 

Finally, one finds  that the following relations are as a whole to hold 

\begin{eqnarray}
&&[{\hat X}_\mu, {\hat X}_\nu ] = - i (\lambda^2/\hbar) {\hat M}_{\mu \nu}
\end{eqnarray}
\begin{eqnarray}
&&[{\hat P} _\mu, (\tau) , {\hat P}_\nu (\tau)] = - i {\hbar} / R(\tau)^2 \ {\hat M}_{\mu \nu}
\end{eqnarray}
\begin{eqnarray}
&&[{\hat X}_\mu, {\hat  P}_\nu (\tau)] = - i \hbar {\hat N} (\tau) {\hat M}_{\mu \nu}
\end{eqnarray}
.\begin{eqnarray}
[{\hat N} (\tau), {\hat X}_\mu ] = - i \lambda^2 /\hbar  {\hat P}_\mu (\tau)
\end{eqnarray}
\begin{eqnarray}
[{\hat N} (\tau), {\hat P}_\mu ] = i \hbar / R(\tau)^2\ {\hat X}_\mu,
\end{eqnarray}
with other familiar relations concerning  ${\hat M}_{\mu \nu}$'s omitted.

\end{document}